\documentclass[aps,PRApplied,twocolumn,superscriptaddress,showpaFcks]{revtex4-1}
\usepackage{amsfonts}
\usepackage{amsmath}
\usepackage{amssymb}
\usepackage{graphicx}
\usepackage{epstopdf}
\usepackage{color}
\usepackage{bbold}
\usepackage{ulem}

\usepackage[colorlinks, citecolor=blue]{hyperref}

\newcommand{\YZ}[1]{{\color{green} #1}}

\renewcommand{\vec}[1]{\boldsymbol{#1}}

\begin{document}

%\title{Arbitrary Multi-Qubit Reset in a Dissipative Quantum Network}

\title{External Control of Qubit-Photon Interaction and Multi-Qubit Reset in a Dissipative Quantum Network}

\author{Xian-Peng Zhang}
\affiliation{Fujian Key Laboratory of Quantum Information and Quantum Optics and Department of Physics, Fuzhou University, Fuzhou 350116, China}
\affiliation{Donostia International Physics Center (DIPC), Manuel de
Lardizabal, 4. 20018, San Sebastian, Spain}
\affiliation{Centro de Fisica de Materiales (CFM-MPC), Centro Mixto CSIC-UPV/EHU,
20018 Donostia-San Sebastian, Basque Country, Spain}

\author{Li-Tuo Shen}
\affiliation{Fujian Key Laboratory of Quantum Information and Quantum Optics and Department of Physics, Fuzhou University, Fuzhou 350116, China}

\author{Yuan Zhang}
\affiliation{Key Laboratory of Material Physics, Ministry of Education, School of Physics and Microelectronics, Zhengzhou University, Zhengzhou 450052, China}

\author{Luyan Sun}
\affiliation{Center for Quantum Information, Institute for Interdisciplinary Information Sciences, Tsinghua University, Beijing, 100084, China}

\author{Huaizhi Wu}
\affiliation{Fujian Key Laboratory of Quantum Information and Quantum Optics and Department of Physics, Fuzhou University, Fuzhou 350116, China}

\author{Zhen-Biao Yang}
\email{zbyang@fzu.edu.cn}
\affiliation{Fujian Key Laboratory of Quantum Information and Quantum Optics and Department of Physics, Fuzhou University, Fuzhou 350116, China}

\author{Zhang-Qi Yin}
\email{zqyin@bit.edu.cn}
\affiliation{Center for Quantum Technology Research and Key Laboratory of Advanced Optoelectronic Quantum Architecture and Measurements (MOE),
School of Physics, Beijing Institute of Technology, Beijing 100081, China}

\begin{abstract}
A quantum network is a promising quantum many-body system because of its tailored geometry and controllable interaction. Here, we propose an external control scheme for the qubit-photon interaction and multiqubit reset in a dissipative quantum network, which comprises superconducting circuit chains with microwave drives and filter-filter couplings. The traditional multiqubit reset of the quantum network requires physically disconnected qubits to prevent their entanglement. However, we use an original effect of dissipation, i.e., consuming the entanglement generated by qubits' interaction, to achieve an external control of the multiqubit reset in an always-connected superconducting circuit. The reset time is independent of the number of qubits in the quantum network. Our proposal can tolerate considerable fluctuations in the system parameters and can be applicable to higher-dimensional quantum networks.
\end{abstract}

\maketitle

\section{Introduction}

The turn of this century has witnessed many advances in the frontier of quantum information processing (QIP), including quantum
computation \cite{nielsen2002quantum,mermin2007quantum}, quantum communication \cite%
{Nature2008-453-1023,Science2007-316-1316}, and quantum simulation \cite%
{NP2012-8-264,NP2012-8-267}. Owing to its tailored geometry and tunable interaction \cite{ritter2012elementary,Xue2016}, quantum network has
attracted considerable attention in cavity quantum electrodynamics (QED) systems \cite{ritter2012elementary,wilk2007single} and electronic circuits \cite{imhof2018topolectrical,song2020realization}. For instance, state-of-the-art superconducting circuits are crucial for the quantum simulation of many-body systems in condensed matter physics \cite{houck2012chip,hartmann2006strongly,schmidt2013circuit, fitzpatrick2017observation}. It is also a compelling platform for realizing topological photonics \cite{wang2016detecting,ozawa2019topological}, topological magnon insulators \cite{cai2019observation}, and high-dimensional topological insulators \cite{wang2020circuit}.

\begin{figure}[t]
\centering
\includegraphics[width=0.98 \columnwidth,angle=0]{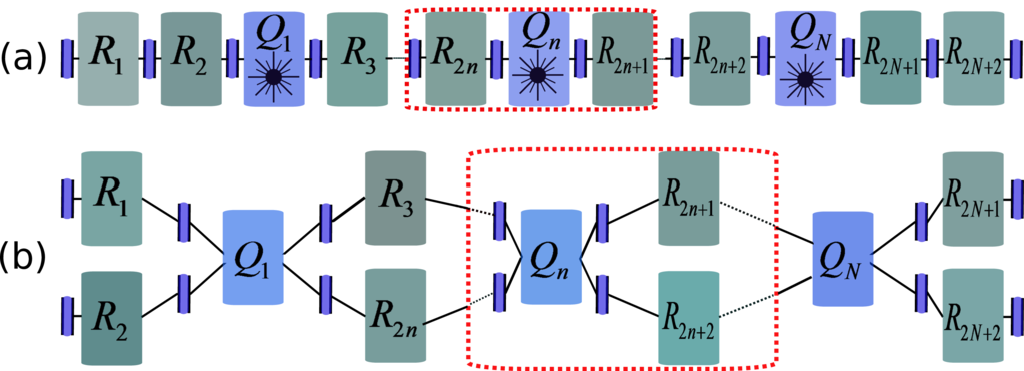}
\includegraphics[width=0.98 \columnwidth,angle=0]{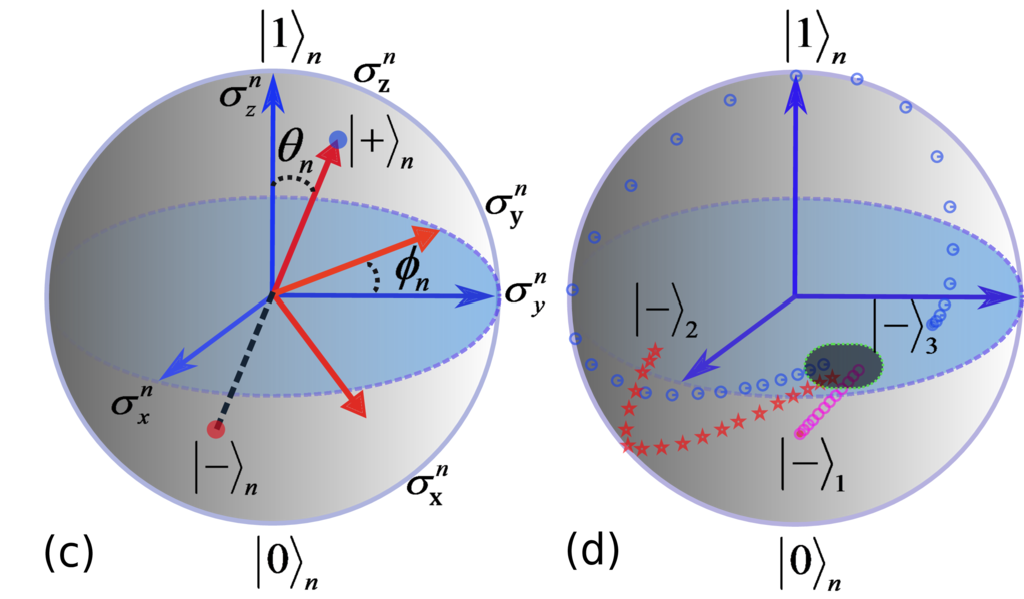}
\caption{Schematic of the multiresonator multiqubit circuit quantum electrodynamics (QED)
architecture where $Q_n$ denotes superconducting qubits (blue rectangular) and $R_n$ denotes
resonators (green rectangular). (a) Cartoon circuit QED architecture of Hamiltonian $H_1$ in Eq. \eqref{H1}. (b) Effective cartoon circuit QED architecture of Hamiltonian $H_{I}$ in Eq. \eqref{HN14}.
(c) Rotation of Pauli operators as shown in Eq. \eqref{Rotation}. (d) A drive of
a low-temperature quantum bath. For more information, see the text. }
\label{f1}
\end{figure}

The conventional route to generate the coherence and the entanglement relies on a demanding control over the operation time and coupling constant. It can be replaced by a relatively open environment, which allows the dissipation to assist the generation of coherence and
entanglement \cite{gong2019verification, basilewitsch2019reservoir, PRA2013-88-023849, N2013-504-419}. One of the typical applications is the cavity- or resonator-assisted qubit reset. The kinetic energy is consumed through
a dissipative environment, e.g., the cavity or the resonator photon loss. Based on the vast convenience and robustness, it has
become attractive in artificial atoms \cite%
{PRL2012-109-183602, PRL2013-110-120501, egger2018pulsed, magnard2018fast},
genuine atoms \cite{NL2004-428-50,PRL2009-103-103001}, spins \cite%
{PRA2010-82-041804,PRL2014-112-050501}, and mechanical objects \cite%
{NL2006-444-71,NL2006-444-67,Yuan2020}. A paradigmatic example is resetting a single qubit through quantum reservoir engineering (QRE), where the dissipation can be engineered to make the system stable toward arbitrary
states (equator of the Bloch sphere) \cite{PRL2012-109-183602}. It was recently demonstrated that resonator-assisted
QRE could be used to prepare the superconducting flux
qubit into any orbital state of the Bloch sphere surface with a controllable
phase factor \cite{PRA2015-91-013825}. The previous research focused on the one-qubit system; however, the multiqubit system has not been investigated.

In multiqubit QIP, the traditional qubit reset protocols require the physically disconnecting qubits to prohibit their entanglement. Hence, it reduces into a one-body problem.
In this study, we investigate the possibility of qubit reset in a many-body category, in which the entanglement among the interacting qubits is inevitable. Unlike the previous schemes \cite{PRA2013-88-023849,N2013-504-419}, where dissipation is used to generate the entanglement, we investigate the dissipation's original effect, i.e., consuming the entanglement generated by the qubits' interaction \cite{breuer2002theory,agarwal1974quantum}. We show how to realize the arbitrary multiqubit reset in an always-connected circuit network. The always-connected circuit network refers to the network where
the neighboring nodes are physically connected with a strong coupling constant. However, whether they are
logically connected is completely determined by the external control (for
instance, the microwave drives applied to each node). Theoretically, the
multiqubit reset in an always-connected circuit network is an extension of the
conception of initialization in the regime of many-body problems. Experimentally, it can make the subsequent multiqubit QIP \cite
{Nature2008-453-1023,Science2013-339-1169} more convenient. Thus, it is unnecessary to switch off the direct or indirect qubits’ interaction during the initialization \cite{PRL2008-101-080502, PRL2015-114-080501}. It might be useful for the future far-ranging quantum device applications based on quantum networks, where the separation of preparation and operation is impractical. Besides, it is essential for further nonlocal operations among different nodes \cite{QIP2016-15-185}.

This study employs the controllable qubits frequencies and microwave drives to realize an external control of the qubit-photon interaction and implement optional qubit reset in such an always-connected circuit network. Independent and precise control over the qubit frequency and microwave drive on each qubit allow us to asynchronously prepare any number of qubits
into arbitrary well-defined initial states. The reset time does not depend on the number of qubits. For initializing the states on the lower Bloch hemisphere, the reset time is in the $0.2-0.8$ $\mu $s range for the experimentally
feasible sample parameters. It is significantly shorter than the superconducting qubit’s intrinsic energy relaxation time in the $6-20$ $\mu $s range \cite{PRB2016-93-104518,PRL2014-113-123601}. Deterministically and quickly initializing qubit into any well-defined state provides a convenient avenue for error-corrected
information processors \cite{S2011-332-1059,N2012-482-382},
and quantum memories \cite%
{PRL2010-105-140501,PRL2010-105-140502}. The reset time becomes longer for the target state moving northwards in the Bloch sphere and finally getting into its limit at the north pole. This limit helps to achieve an optional reset. Finally, the proposed scheme tolerates considerable fluctuation in system parameters. It is also available for 2D and 3D quantum networks.

\section{Model and Theory}

The computational basis states of the proposed model are defined with the two lowest energy states, $\left\vert 0\right\rangle $ and $\left\vert
1\right\rangle $, of the superconducting qubit \cite%
{PRL2011-107-240501}. Fig. \ref{f1}(a) shows that a superconducting qubit ($%
Q_{n}$) is capacitively coupled to two of its adjacent resonators ($R_{2n}$
and $R_{2n+1}$), constructing the building block (red dashed rectangular). $N$ building blocks are capacitively coupled to each other, forming the multiresonator multiqubit circuit QED network. The qubits interact through
two strongly coupled resonators, which serve as a Purcell filter, to suppress the off-resonant interactions \cite{PRL2008-101-080502,PRL2015-114-080501}. In the presence of $N$ independently
adjustable microwave drives, acting one-to-one on $N$ qubits, the near-resonant oscillation
between energy states $\left\vert 0\right\rangle $ and $\left\vert
1\right\rangle $ of each qubit is induced with regulatable frequency $\omega _{n}$.

In rotating frame of $R_{1}=\sum_{m=1}^{2N+2}\omega
_{L}a_{m}^{+}a_{m}+\sum_{n=1}^{N}\omega _{L}\sigma _{z}^{n}/2$, the circuit QED system can be described using the following Hamiltonian (see details in Appendix \ref{SH})
\begin{align} \label{H1}
H_{1} &=\sum_{m=1}^{2N+2}\delta \omega
a_{m}^{+}a_{m}^{{}}+%
v\sum_{n=1}^{N+1}(a_{2n-1}^{{}}a_{2n}^{+}+a_{2n-1}^{+}a_{2n}^{{}}) \\
&+\sum_{n=1}^{N}[g(a_{2n}^{+}+a_{2n+1}^{+})\sigma
_{-}^{n}+g(a_{2n}^{{}}+a_{2n+1}^{{}})\sigma _{+}^{n}]\notag \\
&+ \sum_{n=1}^{N}(\Omega _{n}\sigma _{-}^{n}+\Omega^{*}
_{n}\sigma _{+}^{n}+\delta \varpi _{n}\sigma _{z}^{n}/2),\notag
\end{align}%
with $\delta \omega =\omega _{c}-\omega _{L}$ and $\delta \varpi _{n}=\omega
_{n}-\omega _{L}$. $\sigma _{j}^{n}$ ($j=x,y,z$) and $\sigma _{+}^{n}$
($\sigma _{-}^{n}$) are Pauli operators and raising (lowering)
operators of the $n$-th qubit with the frequency $\omega _{n}$, and $a_{m}$ ($%
a_{m}^{+}$) is the annihilation (creation) operator of the $m$th resonator
with the frequency $\omega _{c}$. $\Omega _{n}$ are
the Rabi frequencies of the drive with frequency $\omega _{L}$ acting on
the $n$-th qubit, $v$ is the filter-filter coupling, and $g$ is the
qubit-filter coupling. For simplicity, we set the reduced plank constant to be one ($\hbar=1$). We made the standard rotating wave
approximation (RWA) to remove any Hamiltonian’s time-dependent terms in
the parameter regime $\omega _{c},\omega _{L},\omega _{n}\gg g,\kappa
,v,\Omega _{n}$.

Figs. \ref{f1}(c) and (d) show the concept of the proposed model. We first rotate the
Pauli operators of each qubit, as shown in Fig. \ref{f1}(c), making
the low-eigenvalue eigenstate of the rotated Pauli operator to be the target
state of the $n$-th qubit. Second, the photon-loss-assisted driving could
stabilize each qubit to its redefined ground state, as shown in Fig. \ref{f1}(d). 

\subsection{Qubit-Photon Interaction under the Rotated Basis}

Let us start with the rotation of the qubit basis and the external control of the qubit-photon interaction in the rotated basis. The target state of the $n$-th qubit $\left\vert -\right\rangle _{n}=\left\vert \Psi (\theta _{n},\phi _{n})\right\rangle _{n}\equiv \cos (%
\frac{\theta _{n}}{2})\left\vert 0\right\rangle_{n} +e^{i\phi _{n}}\sin (\frac{%
\theta _{n}}{2})\left\vert 1\right\rangle_{n} $ can be represented by the points on the Bloch sphere, specified with the angles $\theta _{n}\in \lbrack
0,\pi ]$ and $\phi _{n}\in \lbrack 0,2\pi )$. It is the eigenstate of the Pauli operator after rotation $\sigma _{\mathbf{z}}^{n}=-\sin \theta
_{n}\cos \phi _{n}\sigma^n _{x}+\sin \theta _{n}\sin \phi _{n}\sigma^n _{y}+\cos
\theta _{n}\sigma^n _{z}$ with eigenvalue $\mathcal{E}=-1$. Fig. \ref{f1}(c) shows that this
rotation of Pauli operators, $R^{n}$, can be realized by a rotation around z-axis with angle $%
\phi _{n}$ followed by a rotation around the y-axis with angle $\theta _{n}$
\begin{align} \label{Rotation}
\begin{bmatrix}
\sigma _{\mathbf{x}}^{n} \\
\sigma _{\mathbf{y}}^{n} \\
\sigma _{\mathbf{z}}^{n}%
\end{bmatrix}%
 =
\begin{bmatrix}
\cos \theta _{n}\cos \phi _{n} & \mathtt{-}\cos \theta _{n}\sin \phi _{n} &
\sin \theta _{n} \\
\sin \phi _{n} & \cos \phi _{n} & 0 \\
\mathtt{-}\sin \theta _{n}\cos \phi _{n} & \sin \theta _{n}\sin \phi _{n} &
\cos \theta _{n}%
\end{bmatrix}%
\begin{bmatrix}
\sigma _{x}^{n} \\
\sigma _{y}^{n} \\
\sigma _{z}^{n}%
\end{bmatrix}%
 .
\end{align}%
Here, the bold subscripts $\mathbf{x},\mathbf{y},\mathbf{z}$ denote the space basics after rotation.
The rotation angles $\theta _{n}$ and $\phi _{n}$ determine both the $n$-th Rabi frequency and the detuning of the $n$-th drive field \cite{PRA2015-91-013825} through the relations
\begin{equation} \label{E4N}
-\frac{\mathrm{Re}(\Omega _{n})}{\sin \theta _{n}\cos \phi _{n}}=\frac{%
\mathrm{Im}(\Omega _{n})}{\sin \theta _{n}\sin \phi _{n}}=\frac{\delta
\varpi _{n}}{2\cos \theta _{n}}.
\end{equation}%
We then define the above ratio as an effective Rabi frequency
\begin{equation} \label{ERBF}
\bar{\Omega}\equiv \lbrack |\Omega _{n}|^{2}+|\delta \varpi
_{n}|^{2}/4]^{1/2}.
\end{equation}%
Here, we have removed the $n$-dependence for simplicity.

In rotating frame of $%
R_{2}=v\sum_{n=1}^{N+1}(a_{2n-1}^{{}}a_{2n}^{+}+a_{2n-1}^{+}a_{2n}^{{}})+\sum_{n=1}^{N}\bar{\Omega}\sigma _{\mathbf{z}}^{n}+%
\sum_{m=1}^{2N}\delta \omega a_{m}^{+}a_{m}$, the Hamiltonian \eqref{H1} generates six modes with frequencies $\omega _{lk}=\delta \omega +lv+2k\bar{\Omega} (l=\pm 1,k=0, \pm 1 )$. In this study, we choose the mode with frequency $\omega _{-1,-1}$, whose dynamics prefers the target state $\left\vert -\right\rangle _{n}$ in thermal equilibrium (see more details in Appendix \ref{SH}). By setting $\delta\omega=v+2\bar{\Omega}$, we reach the compact interacting Hamiltonian 
\begin{equation} \label{HN14}
H_{I}=\sum_{n=1}^{N}\sum_{m=2n-1}^{2n+2}g^{m}_{n}a_{m}^{+}\sigma _{-}^{n(\mathbf{z})}e^{i\phi
_{n}}+H.c.\;.
\end{equation}%
Here, $\sigma _{\pm }^{n(\mathbf{z})}=(\sigma _{\mathbf{x}}^{n}\pm i\sigma _{%
\mathbf{y}}^{n})/2$ are the ladder operators in the $\mathbf{z}$-basis.
Again, we have used the RWA in the parameter regime, where $2\bar{\Omega},2v$ are larger than the dissipation rate of the resonator, $\kappa $, and the qubit-filter coupling $g$, i.e., $2\bar{\Omega}%
,2v\gg \kappa ,g$. The effective qubit-photon couplings are given by $[g^{2n-1}_{n},g^{2n}_{n},g^{2n+1}_{n},g^{2n+2}_{n}]=g_n[-1,1,1,-1]$, where $g_n=(g/4)(\cos \theta _{n}+1)$. The later becomes zero, i.e., $g_{n}=0$ for $\theta _{n}=\pi$, which means that the qubit is logically disconnected ($g_{n}=0$, see Fig. \ref{f1}(b)).
Although, it is physically connected with the neighboring ones ($2v\gg \kappa,g$, see Fig. \ref{f1} (a)). Thus, we realize an optional reset. If required, we pick out $\mathcal{N}(<N)$
qubits, and adjust the experimental parameters $\left[ \mathrm{Re}(\Omega
_{n_{i}}),\mathrm{Im}(\Omega _{n_{i}}),\delta \varpi _{n_{i}}/2\right]
=[0,0,-\bar{\Omega}]$ ($g_{n_i}=0$, $i=1,2,\cdots,\mathcal{N}$), as described in Eq. \eqref{E4N}. The vast detuning of
microwave drive stops the dynamic of the $n_{i}$-th qubit. Here, we emphasize that the effective qubit-photon coupling’s external control, referring to the rotated basis, as shown in Eq. \eqref{Rotation}, is different from that of Ref. \cite{zeytinouglu2015microwave} with qubit levels defined in original basis. The effective qubit-photon interaction is independent of the effective Rabi frequency \eqref{ERBF} in the limit of $2\bar{\Omega}\gg \kappa,g$, which aims to make the model $\omega_{-1,-1}$ well-separated from the rest and verify the second RWA.

\subsection{The Multiqubit Reset Driven by Huge Photon Loss}

Next, we investigate the multiqubit reset driven by vast photon loss. Let us start with a qualitative discussion about the multiqubit reset. The interaction Hamiltonian \eqref{HN14} describes the effective circuit network qubits interacting with two shunt-wound
resonators \textit{without} microwave devices and filter-filter coupling, as shown in Fig. \ref{f1}(b). The dynamics of this effective
circuit network is not difficult to guess. Although the indirect qubit-qubit
interaction realized through qubit-filter coupling ($g_{n}\neq 0$), generates correlation (entanglement) among
qubits. It will be easily consumed by dissipation caused by the vast photon
loss of resonators (see figures and discussions at the end of Appendix \ref{dmme}). All qubits are driven into their rotated ground states $%
\left\vert -\right\rangle_{n}$ by QRE, as shown in Fig. \ref{f1}(d). The rotated ground states, i.e., the target states, are
controlled by the external microwave drives and the frequencies of the
corresponding qubits, i.e., $\left[\mathrm{Re}(\Omega _{n}), \mathrm{Im}%
(\Omega_{n}),\delta\varpi_{n}/2\right]$. The unavoidable fluctuation of the
above parameters redefines the rotated ground state and slightly affects the
target state \cite{PRA2015-91-013825}. Under the condition, the two RWAs are valid, other parameters, such as $g$, $\kappa$, $v$, merely
influence the reset time and allow quite a large region of change.

To obtain a quantitative expression of reset time, we solve
the Markovian master equation. The Lindblad master equation can model the evolution of the multiresonator multiqubit circuit in an open environment.
\begin{equation} \label{ELME}
\frac{d}{dt}\rho (t)=L[H_{I}]\rho (t)+D_{c}\rho (t),
\end{equation}%
with
\begin{equation}
D_{c}=\sum_{m=1}^{2N+2}\frac{\kappa }{2}((1+\bar{n})D[a_{m}]+\bar{n}%
D[a_{m}^{+}]).
\end{equation}%
$L$ is the superoperator $L[H_{I}]\rho (t)=-i[H_{I},\rho (t)]$, describing the
unitary evolution under the domination of $H_{I}$.
$D_{c}$ is a dissipator,
representing the dissipative environment created by resonator photon loss, $%
D[O]\rho =2O\rho O^{+}-\{O^{+}O,\rho \}$. $\bar{n}=1/\left( e^{\omega _{c}/k_{B}T_{c}}-1\right) $ is the photon number operator’s expectation value in equilibrium for resonator with temperature, $T_{c}$, with $k_{B}$ being Boltzmann constant, where we have not considered the feedback effect of superconducting qubits on the resonators. For simplicity, we assumed the same $\bar{n}$ and $\kappa$ for each resonator.

The state population of each qubit is defined as $P^n_{\mathcal{E}}(t)=\textrm{Tr}_{q} \textrm{Tr}_{r} \{\rho(t) \mathcal{P}_n(\mathcal{E})\}$ ($\mathcal{E}=\pm 1$), where $\mathcal{P}_n(\mathcal{E})=\vert \mathcal{E}\rangle_n
\langle \mathcal{E}\vert $ is the projection operator of the $n$-th qubit and $\textrm{Tr}_{q}$ ($\textrm{Tr}_{r}$) mean trace over the multiqubit (multiresonator) space, respectively.
With the definition of $\vec{P}^{n}_{\mathcal{E}}(t)=[P^n_{-1}(t),P^n_{+1}(t)]^{T}$, the Lindblad master equation \eqref{ELME} will reduce to a rate equation for the state population (see Appendix \ref{dmme} for details)
\begin{equation} \label{E12}
\frac{d}{dt}\vec{P}^{n}(t)=-\Gamma _{n}\vec{N}\vec{P
}^{n}(t),
\end{equation}%
with
\begin{equation}
\mathbf{N}=\left[
\begin{array}{cc}
\bar{n} & -(\bar{n}+1) \\
-\bar{n} & +(\bar{n}+1)%
\end{array}%
\right]\;.
\end{equation}%
Here, $\Gamma _{n}=16g_n^{2}/\kappa$ is the effective polarization rate of each qubit. Although the nonlocal qubit-qubit correlation (or entanglement) can be formed due to the indirect qubit-qubit interaction realized by photon-mediated qubit-photon coupling, it will be easily consumed by dissipation caused by vast photon loss of resonators ($\kappa \gg g_n$), as discussed in Appendix \ref{dmme}.
It is worth knowing that the state population of superconducting qubit satisfies $\partial _{t}\vec{P}^{n}(t)=0$ in a steady state. We obtain the expectation value of the operator $\sigma _{\mathbf{z}}^{n}$ for the equilibrium state
\begin{equation}
\left\langle \sigma _{\mathbf{z}}^{n}\right\rangle _{eq}=\frac{e^{-\omega
_{c}/k_{B}T_{c}}-1}{e^{-\omega _{c}/k_{B}T_{c}}+1}.
\end{equation}%
In an ideal case where all resonators are cooled to their ground states,
i.e., vacuum states ($T_{c}\rightarrow 0$), the final expectation value is
approximately $\left\langle \sigma _{\mathbf{z}}^{n}\right\rangle
_{eq}\simeq -1$.

\section{Results and Discussions}

\begin{figure}[t]
\centering
\includegraphics[width=0.98 \columnwidth,angle=0]{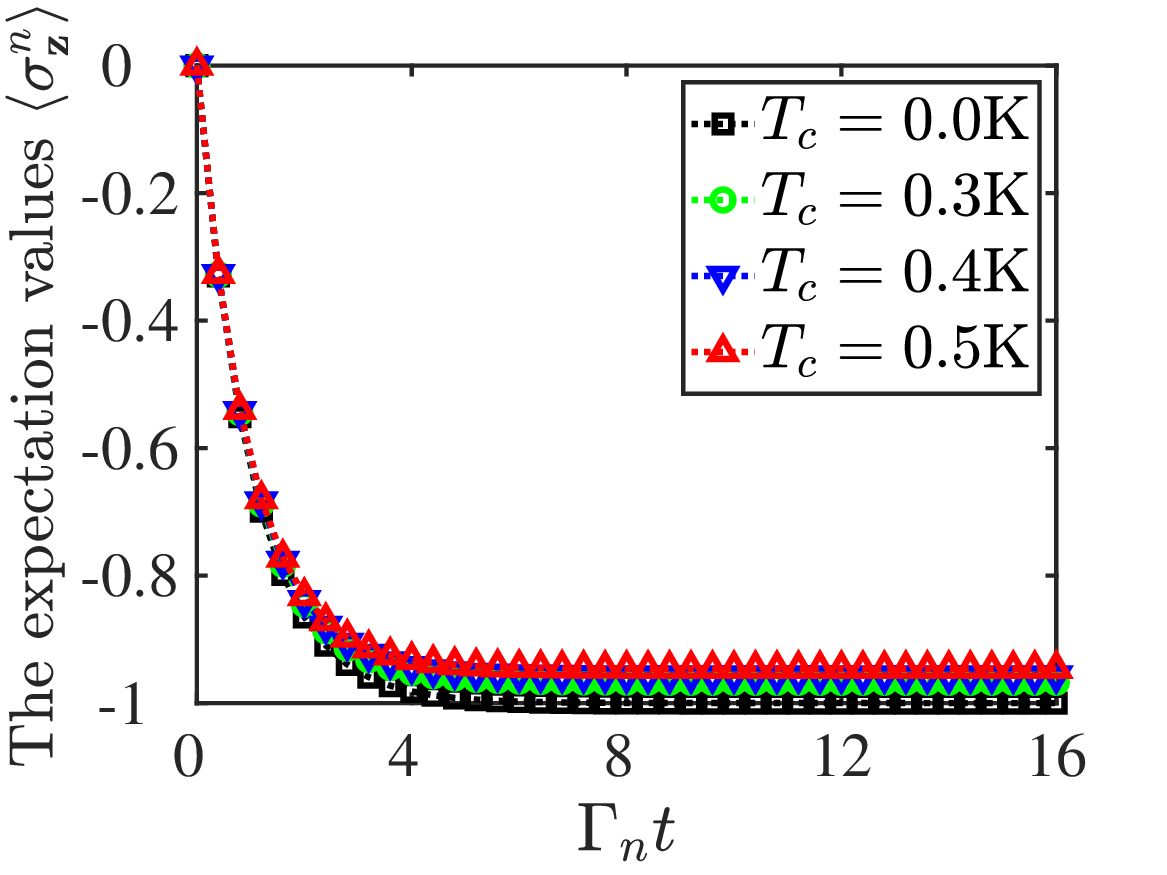}
\caption{The evolution of the simulated expectation values $\left\langle
\protect\sigma^n_\textbf{z} \right\rangle$ for different temperatures $%
T_c$. Other parameters are $v/2\protect\pi=100$ MHz \cite{PRL2015-114-080501}, $\bar{\Omega}/2\protect%
\pi=100$ MHz \cite{braumuller2017analog}, $\protect\omega_L/2\protect\pi=5.7$ GHz, $\protect\omega_c/2%
\protect\pi=6$ GHz \cite{PRL2008-101-080502,PRL2015-114-080501}, $\Omega _{n}=\bar{\Omega}e^{i(\protect\pi-\protect\phi%
_n)}$ and $\protect\omega_n=\protect\omega_L+2\bar{\Omega}\cos \protect%
\theta _{n}$ \cite{PRL2015-114-080501,steffen2006state} \YZ{.}}
\label{f2}
\end{figure}

The separability of the rate equation (\ref{E12}) for each qubit makes it easy to simulate the gratifying results. For the $n$-th qubit, initially assumed to be maximally mixed in the basis ($P^n_{\pm 1}(0)=1/2$), the time evolution of the
simulated expectation values $\left\langle \sigma _{\mathbf{z}%
}^{n}\right\rangle $ is shown in Fig. \ref{f2}. The equilibrium expectation values for different temperatures $T_{c}=(0.0,0.3,0.4,0.5)$K are $%
-1.000,-0.967,-0.957,-0.946$, respectively. For temperature $T_{c}\leq 0.3$K (where the thermodynamics effect can be ignored \cite{PRB2016-93-104518,PRL2014-113-123601}), they can be fitted to an
exponential function to obtain an effective resetting time, $%
T_{n}$. A fit with $\left\langle \sigma _{\mathbf{z}%
}^{n}\right\rangle =\textrm{exp}(-t/T_{n})-1$ yields
\begin{equation}
T_{n}\simeq \frac{1}{\Gamma _{n}}=\frac{\kappa}{16g_n^{2}}=\frac{\kappa}{(1+\cos
\theta _{n})^{2}g^{2}}. \label{E11}
\end{equation}%
We obtained that the reset time was independent of the number of
qubits (see detailed discussions at the end of Appendix \ref{dmme}). Additionally, the reset time becomes longer for the target
state moving northwards. The most efficient reset occurs in $\hat{z}$ direction with the effective dissipation rate $\Gamma _{n}=4g^{2}/\kappa $ ($\theta _{n}=0$, $|-\rangle_{n}=|0\rangle _{n}$). For the worst case, the effective dissipation rate approaches zero, i.e., $%
\Gamma _{n}\rightarrow 0$, corresponding to the reset happening
around $-\hat{z}$ direction ($\theta _{n}=\pi $, $|-\rangle _{n}=|1\rangle
_{n}$).

In the following, we present the setting of the experimental parameters. If the $n$-th qubit should be prepared to state $\left\vert
-\right\rangle _{n}=\left\vert \Psi (\theta _{n},\phi _{n})\right\rangle _{n}$, its
qubit frequency should be set to be $\omega_n=\omega_L+2\bar{\Omega}\cos
\theta _{n}$ \cite{PRL2015-114-080501,steffen2006state}, and the microwave drive acting on this qubit should be
adjusted with $\left[\mathrm{Re}(\Omega _{n}), \mathrm{Im}(\Omega
_{n}),\delta\varpi_n/2\right]=\left[-\bar{\Omega}\sin \theta _{n}\cos \phi
_{n},\bar{\Omega}\sin \theta _{n}\sin \phi _{n},\bar{\Omega}\cos \theta _{n}%
\right]$. In addition, the frequency of cavity mode should be tuned to satisfy $\omega_c=\omega_L+ v+2\bar{\Omega}$. In the parameter regime $(g,\kappa)/2\pi=(2,20)$ MHz \cite{PRL2008-101-080502}, the reset time is between
the range of $0.2-0.8\mu$s for lower Bloch hemisphere, which is significantly
shorter than the intrinsic energy relaxation time for the superconducting
flux qubit in the 6-20$\mu$s range \cite%
{PRB2016-93-104518, PRL2014-113-123601}. The resetting time becomes longer
for the target state with a larger $\theta_n$ angle, which tends to the
limit at the north pole.

\begin{figure*}[t]
\centering
\includegraphics[width=1.00\textwidth]{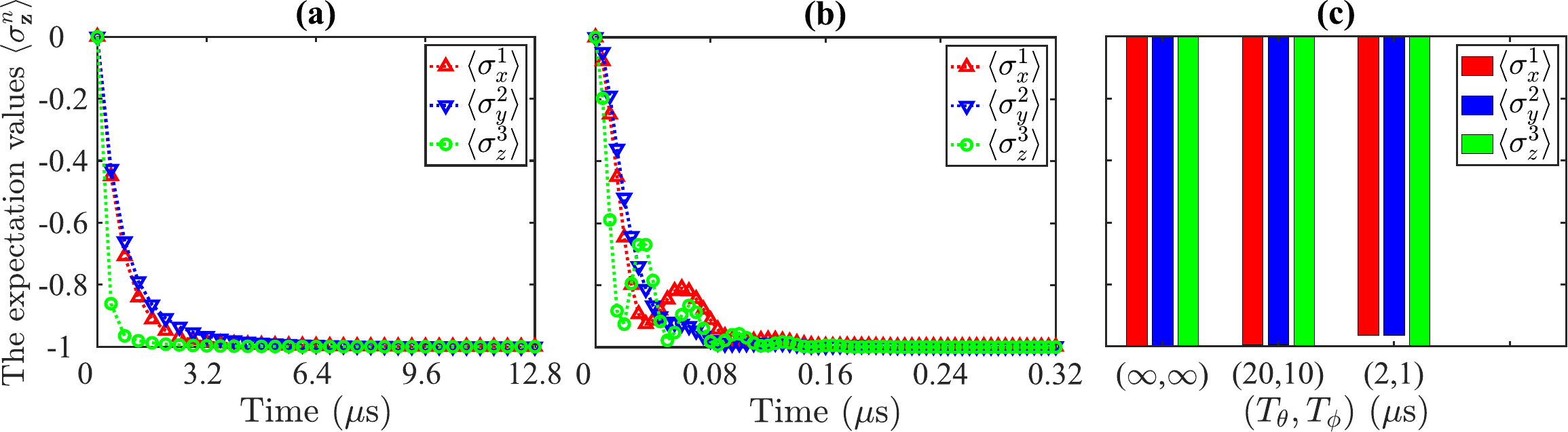} %
\caption{(a)-(b) The evolution of the simulated expectation values $\left\langle \protect\sigma^n_\mathbf{z} \right\rangle$ for the parameters (a) $(g,%
\protect\kappa)/2\protect\pi=(2,20)$ MHz \cite{PRL2008-101-080502} and (b) $(g,\protect\kappa)/2\protect\pi%
=(15,10)$ MHz \cite{PRL2008-101-080502}, respectively. (c) The final expectation values $\left\langle
\protect\sigma^n_\mathbf{z} \right\rangle$ for the parameters $(g,\protect\kappa)/2\protect\pi=(15,10)$
MHz for the three different sets of qubit dissipation rates ($1/T_\protect%
\protect\theta$, $1/T_\protect\protect\phi$) \cite{PRL2008-101-080502, PRL2014-113-123601, schreier2008suppressing}. Other parameters are the same as shown in Fig. \ref{f2}, and the temperature is zero, i.e., $T_c=0$K. }
\label{f3}
\end{figure*}

For the numerical results, a Monte Carlo method is used to simulate the
Lindblad master equation of the interaction Hamiltonian $H_1$ assisted by
Quantum Toolbox in Python \cite{CPC2013-184-1234}. Assume our system has three qubits, which should be reset into $\left[\left\langle\sigma_x^{1}\right\rangle,\left\langle%
\sigma_y^{2}\right\rangle,\left\langle\sigma_z^{3}\right\rangle\right]%
=[-1,-1,-1]$ state. Fig. \ref{f3}(a) describes the evolution of the expectation values $\left\langle
\sigma^n_\mathbf{z} \right\rangle$ for Lindblad master equation. All qubits
are able to reach a quite high reliability with the final expectation values
$\left[\left\langle\sigma_x^{1}\right\rangle,\left\langle\sigma_y^{2}\right%
\rangle,\left\langle\sigma_z^{3}\right\rangle \right]\approx[%
-0.9998,-0.9997,-1.0000] $. We observed that the decay of the
simulated expectation value $\left\langle \sigma^3_z \right\rangle$
(approaching $-1$) was four times faster than those of the simulated
expectation values $\left\langle \sigma^1_x \right\rangle$ and $\left\langle
\sigma^2_y \right\rangle$, validating the analytical derivation of the
effective $\theta_n$-dependent resetting time (\ref{E11}).
The influences of the fluctuations of parameters $\Omega _{n}$ and $%
\delta\varpi_n$ have been previously studied and are sufficiently small to be neglected \cite{PRA2015-91-013825}. Simulation results show that considerable fluctuations of parameters $\kappa$ and $g$ are also allowed.
The reset efficiency can be improved when we optimize
the parameters $(g,\kappa)/2\pi=(15,10)$ MHz \cite{PRL2008-101-080502}. As shown in Fig. \ref{f3}(b),
all qubits are almost completely driven into their
target states after 0.32$\mu$s, with equilibrium states $\left[%
\left\langle\sigma_x^{1}\right\rangle,\left\langle\sigma_y^{2}\right\rangle,%
\left\langle\sigma_z^{3}\right\rangle\right] \approx\left[%
-0.998,-0.998,-1.000\right]$.

The qubit dissipation will cause a disturbance in the equilibrium state and
should be considered. In the presence of
qubit decay and qubit dephasing, the Lindblad master equation of the quantum
system is
\begin{align} \label{E16}
\frac{d}{dt}\rho (t) &=L[H_{1}]\rho (t)+D_{c}\rho (t) \\
&+ \sum^N_{n=1} \frac{1}{T_\theta}D[\sigma^{n(z)}_-] \rho (t)+ \frac{1}{%
2T_\phi}D[\sigma^{n(z)}_z]\rho (t), \notag
\end{align}
where $1/T_\theta$ and $1/T_\phi$ are the rates for the qubit decay and qubit
dephasing, respectively. The evolution of the expectation values $\langle
\sigma^n_\mathbf{z} \rangle$, are simulated with the experimentally available parameters $%
(T_\theta,T_\phi)=(20,10)\mu$s \cite{PRL2014-113-123601}. We observe that relatively high
reliability can be obtained with the final expectation values $\left[%
\left\langle\sigma_x^{1}\right\rangle,\left\langle\sigma_y^{2}\right\rangle,%
\left\langle\sigma_z^{3}\right\rangle\right] \approx\left[%
-0.995,-0.995,-1.000\right]$, as shown in Fig. \ref{f3}(c). Thus, in principle,
the proposed scheme is feasible with the presently experimental sample parameters.

\section{Conclusion and outlook}
In conclusion, we demonstrated the external control of the effective qubit-photon interaction in a quantum network comprising superconducting circuit chains with microwave drives and filter-filter couplings.
The tailored geometry and controlled interaction of a quantum network offer a new possibility of realizing topological transition in photon and qubit systems. We used the original effect of dissipation, i.e., consuming the entanglement generated by qubits’ interaction, to realize the arbitrary reset in an always-connected circuit network through QRE. We employed the controllable microwave drives to realize an optional reset in such an always-connected circuit network. Moreover, rapid and independent control over the internal states of superconducting qubits allows us to achieve flexible reset for any designated qubits.
 The proposed result can be easily generalized to 2D and 3D quantum networks.

Finally, let us discuss the potential applications of this quantum network. We note that each node's microwave drives can independently and externally control effective qubit-photon interactions. Reducing the photon (qubit) degrees of freedoms provides independent and external adjustments of the photon-mediated qubit-qubit \cite{van2013photon,PRL2015-114-080501} (qubit-mediated photon-photon \cite{hartmann2006strongly,greentree2006quantum}) interactions. Note that the topologically trivial and nontrivial phases require different photon-photon (qubit-qubit) coupling configurations, necessitating a full control of the effective couplings among neighboring (resonators) qubits. This quantum network provides the new possibility of realizing a topological transition in photon and qubit systems.

\section*{Acknowledgment}
We thank Geza Giedke for useful discussions.
This work was supported by the Natural Science Foundation of China
under Grants No. 11875108, No.11405031, No.11347114, and the Natural Science Foundation of Fujian Province under Grant No.2018J01412, No.2014J05005. Z.-Q.Y. was supported by National Natural Science Foundation of China under Grant No. 61771278 and Beijing Institute of Technology Research Fund Program for Young Scholars. L.-Y.S. was supported by the National Key Research and Development Program of China under Grant No. 2017YFA0304303 and the Natural
Science Foundation of China under Grant No.11925404.

\appendix

\section{System Hamiltonian} \label{SH}

In this section, we derive the interaction Hamiltonian $H_I$. In the presence of $N$ independently
adjustable microwave drives, acting one-to-one on $N$ qubits, an oscillation
between energy levels $\left\vert 0\right\rangle $ and $\left\vert
1\right\rangle $ of each qubit with regulable frequency $\omega _{n}$ is
induced near resonance. As shown in Fig. \ref{f1}(a) of the main text, superconducting qubits ($%
Q_{n}$) are capacitively coupled to each other using two resonators. The composite Hamiltonian of the whole circuit
network is
\begin{align}
  H=H_{0}+H_{r}+H_{d}+H_{h},
\end{align}
with
\begin{align}
H_{0} &=\sum_{m=1}^{2N+2}\omega _{c}a_{m}^{+}a_{m}+\sum_{n=1}^{N}\frac{%
\omega _{n}}{2}\sigma _{z}^{n}, \\
H_{r} &=\sum^{N}_{n=1}ga_{2n+1}^{{}}\sigma _{x}^{n}+ga_{2n}^{{}}\sigma _{x}^{n}+H.c., \\
H_{d} &=\sum_{n=1}^{N}\Omega _{n}\sigma _{-}^{n}e^{i\omega _{L}t}+\tilde{%
\Omega}\sigma _{-}^{n}e^{-i\omega _{L}t}+H.c., \\
H_{h} &=\sum^{N+1}_{n=1}va_{2n-1}^{{}}a_{2n}^{+}+H.c.,
\end{align}%
where $\sigma _{j}^{n}$ ($j=x,y,z$) and $\sigma _{+}^{n}$ ($\sigma _{-}^{n}$%
) are the $j$-direction and raising (lowering) Pauli operators of the $n$-th
qubit with the frequency $\omega _{n}$, and $a_{m}$ ($a_{m}^{+}$) is the
annihilation (creation) operator of the $m$th resonator with the frequency $%
\omega _{c}$. $\Omega _{n}$ and $\tilde{\Omega}_{n}$ are the Rabi and the
counter-rotating Rabi frequencies of the microwave drive with frequency $\omega _{L}$ acting on the $n$-th qubit, $v$
is the filter-filter coupling between resonators, and $g$ is the qubit-filter coupling. $\hbar =1$ is assumed for simplicity. In
rotating frame defined by $R_{1}=\sum_{m=1}^{2N+2}\omega
_{L}a_{m}^{+}a_{m}+\sum_{n=1}^{N}\omega _{L}\sigma _{z}^{n}/2$, the system
Hamiltonian becomes
\begin{align}
H_{1} &=\sum_{m=1}^{2N+2}\delta \omega
a_{m}^{+}a_{m}^{{}}+%
H_h \\
&+\sum_{n=1}^{N}\mathrm{Re}(\Omega _{n})\sigma _{x}^{n}+\mathrm{Im}(\Omega
_{n})\sigma _{y}^{n}+\delta \varpi _{n}\sigma _{z}^{n}/2 \notag \\
&+\sum^{N}_{n=1 }g(a_{2n}^{+}+a_{2n+1}^{+})\sigma
_{-}^{n}+g(a_{2n}^{{}}+a_{2n+1}^{{}})\sigma _{+}^{n},\notag
\end{align}%
with $\delta \omega =\omega _{c}-\omega _{L}$ and $\delta \varpi _{n}=\omega
_{n}-\omega _{L}$. Here, we have made the standard rotating wave
approximation (RWA) to remove any Hamiltonian's time-dependent terms in
the parameter regime $\omega _{c},\omega _{L},\omega _{n}\gg gn,\kappa
,v,\Omega _{n},\tilde{\Omega}_{n}$.

In analogy to resonator-assisted quantum bath engineering \cite%
{PRA2015-91-013825}, we introduce a rotating transformation $\mathrm{R}%
^{\mathrm{n}}$ of Pauli operators for each qubit to investigate the
arbitrary direction's polarization efficiency
\begin{equation}
\begin{bmatrix}
\sigma _{\mathbf{x}}^{n} \\
\sigma _{\mathbf{y}}^{n} \\
\sigma _{\mathbf{z}}^{n}%
\end{bmatrix}%
 =
\begin{bmatrix}
\cos \theta _{n}\cos \phi _{n} & \mathtt{-}\cos \theta _{n}\sin \phi _{n} &
\sin \theta _{n} \\
\sin \phi _{n} & \cos \phi _{n} & 0 \\
\mathtt{-}\sin \theta _{n}\cos \phi _{n} & \sin \theta _{n}\sin \phi _{n} &
\cos \theta _{n}%
\end{bmatrix}%
\begin{bmatrix}
\sigma _{x}^{n} \\
\sigma _{y}^{n} \\
\sigma _{z}^{n}%
\end{bmatrix}%
.
\end{equation}%
We define this ratio as an effective Rabi frequency
\begin{equation} \label{ERBFA}
\bar{\Omega}\equiv \lbrack |\Omega _{n}|^{2}+|\delta \varpi
_{n}|^{2}/4]^{1/2}.
\end{equation}%
Here, we have removed the $n$-dependence by adjusting $\Omega _{n}$
and $\delta \varpi _{n}$ for simplicity. In rotating frame of
\begin{align}
  R_{2}=H_{h}+\sum_{n=1}^{N}\bar{\Omega}\sigma _{\mathbf{z}}^{n}+%
\sum_{m=1}^{MN}\delta \omega a_{m}^{+}a_{m}
\end{align}
\begin{equation}
H_{2}(t)=\sum_{n=1}^{N}\sum_{l=\pm 1}\sum_{k=0,\pm 1}H_{lk}^{n}(t)+H.c.,
\label{a8}
\end{equation}%
with
\begin{align}
H_{lk}^{n}(t)=\left\{
\begin{array}{ll}
\sum^{MN}_{m=1}c_{mn}^{l}\Theta_{-}^{n}g^m_ne^{i\omega _{lk}t}a_{m}^{+}\sigma
_{-}^{n(\mathbf{z})}, & \hbox{k=$-$1;} \\
\sum^{MN}_{m=1}c_{mn}^{l}\Theta_{\mathbf{z}}^{n}g^m_ne^{i\omega
_{lk}t}a_{m}^{+}\sigma _{z}^{n(\mathbf{z})}, & \hbox{k=0;} \\
\sum^{MN}_{m=1}c_{mn}^{l}\Theta_{+}^{n}g_n^me^{i\omega _{lk}t}a_{m}^{+}\sigma
_{+}^{n(\mathbf{z})}, & \hbox{k=+1,}%
\end{array}%
\right. \label{A15}
\end{align}%
\begin{equation}
\Theta_{\pm }^{n}=\frac{1}{2}(\cos \theta _{n}\mp 1)e^{i\phi _{n}},\Theta_{\mathbf{z}%
}^{n}=\frac{1}{2}\sin \theta _{n}e^{i\phi _{n}}, \label{A17}
\end{equation}%
where $\sigma _{\pm }^{n(\mathbf{z})}=(\sigma _{\mathbf{x}}^{n}\pm i\sigma _{%
\mathbf{y}}^{n})/2$ are the ladder operators in the $\mathbf{z}$-basis. The
coefficients $c_{mn}^{l}$, which are independent of the number of qubits $N$, are given in TABLE \ref{table1} for the case $N=3$. The rotating frame of $%
R_{2}$ makes the Hamiltonian generate six modes \textrm{M}$(l,k)(l=\pm
1,k=0,\pm 1)$ with frequencies $\omega _{lk}=\delta \omega +lv+2k\bar{\Omega}
$. We make a brief summation of the functions of these six modes. There
is no preference in the $\sigma _{\mathbf{z}}^{n}$ direction for the
dynamics of modes \textrm{M}$(l,0)$ at the thermal equilibrium, while those
of modes \textrm{M}$(l,\pm 1)$ would drive the qubit to the $\left\langle
\sigma _{\mathbf{z}}^{n}\right\rangle =\pm 1$ states, respectively \cite%
{PRL2014-112-050501, PRA2015-91-013825}. Therefore, modes \textrm{M}$(l,-1)$
must dominate our polarization process. Here, we prefer the mode \textrm{M}$%
(-1,-1)$. We may set $\Delta =\delta \omega -v-2\bar{\Omega}$ to
be close to zero, choose the strong enough filter-filter coupling and the effective Rabi frequency ($2\bar{\Omega},2v\gg \Delta,g,\kappa $), so that
other high-frequency modes \textrm{M}$(l,k)$ are separated from mode \textrm{M}$(-1,-1)$. Therefore, the Hamiltonian (\ref{A15}) reduces to $%
H_{I}(t)=\sum_{n=1}^{N}H_{n}(t)$ with
\begin{equation}
H_{n}(t)=\sum^{2n+2}_{m=2n-1}g^{m}_{n}e^{i(\Delta t+\phi
_{n})}a_{m}^{+}\sigma _{-}^{n(\mathbf{z})}+H.c.. \label{H14}
\end{equation}%
Here, we used RWA in the parameter regime
where $2\bar{\Omega},2v$ are larger than the dissipation rate of the
resonator, $\kappa $, and the qubit-filter coupling $g$, i.e., $2\bar{\Omega}%
,2v\gg \kappa,g$.
The effective filter-qubit couplings for
the dominated mode M($-1,-1$) is given by $[g^{2n-1}_{n},g^{2n}_{n},g^{2n+1}_{n},g^{2n+2}_{n}]=g_n[-1,1,1,-1]$, where $g_n=(g/4)(\cos \theta _{n}+1)$. Thus, $g_{n}=0(\theta
_{n}=\pi )$ in an effective Hamiltonian (\ref{H14}) can be used to realize
asynchronous resetting. If required, we randomly select $\mathcal{N}(<N)$
qubits, and adjust the experimental parameters $\left[ \mathrm{Re}(\Omega
_{n_{i}}),\mathrm{Im}(\Omega _{n_{i}}),\delta \varpi _{n_{i}}/2\right]
=[0,0,-\bar{\Omega}]$ ($i=(1,2,\cdots,\mathcal{N})$). They are
physically connected with neighboring qubits; however, they are logically
disconnected from Hamiltonian (\ref{H14}). The vast detuning of
microwave drive stops the dynamic of the $n_{i}^{th}$ qubit.

\begin{table}[htbp]
\begin{center}
\begin{tabular}{ccccccccc}
\hline\hline
$c_{mn}^{\pm}$ & $m$=1 & $m$=2 & $m$=3 & $m$=4 & $m$=5 & $m$=6 & $m$=7 & $m$%
=8 \\ \hline
$n=1$ & $\pm \frac{1}{2}$ & $\frac{1}{2}$ & $\frac{1}{2}$ & $\pm \frac{1}{2}$
& $0$ & $0$ & $0$ & $0$ \\
$n=2$ & $0$ & $0$ & $\pm \frac{1}{2}$ & $\frac{1}{2}$ & $\frac{1}{2}$ &$\pm \frac{1}{2}$ & $%
0$ & $0$ \\
$n=3$ & $0$ & $0$ & $0$ & $0$ & $\pm \frac{1}{2}$ & $\frac{1}{2}$ & $\frac{1%
}{2}$ & $\pm \frac{1}{2}$ \\ \hline\hline
\end{tabular}%
\end{center}
\caption{The constants $c_{mn}^{l}(l=\pm 1)$ specified in Eq. (\protect\ref%
{A15}) versus the qubit $n$ and resonator indices $m$ for the
system dominated by the mode M$(l,k)$. Here, $N$ and $M$ are assumed to be 3 and 2, respectively.}
\label{table1}
\end{table}

\section{Derivation of Markovian Master Equation} \label{dmme}
The evolution of the multiresonator and multiqubit network for the
superconducting circuit in an open environment can be modeled by the
Lindblad master equation \cite{breuer2002theory},
\begin{equation}
\frac{d}{dt}\rho (t)=L\left[ H_{I}(t)\right] \rho +D_{c}\rho (t),
\label{B1}
\end{equation}%
where $L\left[ H_{I}(t)\right] \rho =-i[H_{I}(t),\rho ]$ is a superoperator
describing the unitary evolution dominated by $H_{I}$, $D_{c}$
is a dissipator describing the non-Hermitian dynamics of the system due to
the coupling to environment
\begin{equation}
D_{c}=\sum_{m=1}^{2N+2}\frac{\kappa }{2}((1+\bar{n})D[a_{m}]+\bar{n}%
D[a_{m}^{+}]), \label{B2}
\end{equation}%
where $D[O]\rho =2O\rho O^{+}-\{O^{+}O,\rho \}$, $\bar{n}=1/\left( e^{\omega _{c}/k_{B}T_{c}}-1\right) $
is the expectation value of the photon number operator at equilibrium
associated with the temperature of the bath, $T_{c}$, and Boltzmann
constant $k_{B}$. For simplicity, we do not consider the feedback effect of transmon qubits on the resonators and assumed the same $\bar{n}$ and $\kappa$ for each resonator.

In the rotating frame with
the dissipator $D_c$, the interaction superoperator can be expressed as
\begin{equation}
\tilde{Q}(t)=e^{-D_{c}t}Q(t)e^{D_{c}t},
\end{equation}
and the system state evolves according to
\begin{equation}
\tilde{\rho}
(t)=e^{-D_{c}t}\rho (t)
\end{equation}
Thus, the master equation (\ref{B1}) of the whole
system reduces to
\begin{equation}
\dot{\tilde{\rho}}(t)=\tilde{L}[H_{I}(t)]\tilde{\rho}(t).
\end{equation}%
Let us introduce a projection operator $\hat{P}$ to satisfy:
\begin{equation}
\hat{P}\rho (t)=\varrho (t)\otimes \rho _{eq}.
\end{equation}%
$\rho _{eq}$ is the resonators equilibrium
state. $\varrho (t)=\mathrm{Tr}_{r}[\rho (t)]$ is the reduced state of $N$
flux-tunable transmon qubits, where $\textrm{Tr}_{r}=\textrm{tr}^1_{r}\cdots \textrm{tr}^{2N+2}_{r}$ means trace over the multiresonator space. The reduced dynamics of the flux-tunable transmon qubits are
transformed into the 2nd order time-convolutionless (TCL) master equation
\cite{breuer2002theory}
\begin{equation}
\frac{d}{dt}\hat{P}\tilde{\rho}(t)=\int_{0}^{t}d\tau \hat{P}\tilde{L}%
[H_{I}(t)]\tilde{L}[H_{I}(t-\tau )]\hat{P}\tilde{\rho}(t).
\end{equation}%

Before we continue the above equation, let us introduce some important properties of dissipator $D_{c}$:
\begin{align} \label{DC1}
D_{c}^{+}[\mathbb{1}] &=\mathbb{0},e^{D_{c}^{+}t}[\mathbb{1}]=\mathbb{1},
\end{align}%
\begin{align} \label{DCA}
D_{c}^{+}[a_{m}] &=-\frac{\kappa }{2}a_{m},e^{D_{c}^{+}t}[a_{m}]=e^{-\kappa
t/2}a_{m},
\end{align}%
\begin{align} \label{DCAP}
D_{c}^{+}[a_{m}^{+}] &=-\frac{\kappa }{2}%
a_{m}^{+},e^{D_{c}^{+}t}[a_{m}^{+}]=e^{-\kappa t/2}a_{m}^{+}.
\end{align}%
The algebraic transformation of the dissipator $D_{c}$ also satisfies
\begin{equation} \label{trcdoo}
\mathrm{Tr}
_{r}\{D_{c}^{+}[O]O^\prime\}=\mathrm{Tr}_{r}\{OD_{c}[O^\prime]\},
\end{equation}
 for the arbitrary
resonator operator $O$ ($O^\prime$). Hence, we have
\begin{align}
\hat{P}\tilde{\rho} =\mathrm{Tr}_{r}\{\mathbb{1}e^{tD_{c}}[\rho] \}\otimes \rho
_{eq} =\mathrm{Tr}_{r}[e^{-tD_{c}^{+}}[\mathbb{1}]\rho ]\otimes \rho _{eq} =\hat{P}\rho.
\end{align}%
Here, we have used the relations \eqref{trcdoo} and \eqref{DC1}.
Using Eq. \eqref{trcdoo},
we obtain the equation of motion of density matrix of multiqubit \cite%
{PRL2014-112-050501}
\begin{align} \label{B11}
\dot{\varrho}(t) &=\int_{0}^{t}d\tau \mathrm{Tr}_{r}\{L[H_{I}(t)]e^{
D_{c}\tau}L[H_{I}(t\mathtt{-}\tau )]\varrho (t)\otimes \rho _{eq}\}
 \notag \\
&=\int_{0}^{t}d\tau \mathrm{Tr}_{r}\{e^{
D_{c}^{+}\tau}(L[H_{I}(t)])L[H_{I}(t\mathtt{-}\tau )]\varrho
(t)\otimes \rho _{eq}\} \notag \\
&=\int_{0}^{t}d\tau e^{-\kappa \tau /2}\mathrm{Tr}%
_{r}\{L[H_{I}(t)]L[H_{I}(t\mathtt{-}\tau )]\varrho (t)\otimes \rho
_{eq}\} \notag \\
&=\mathtt{-}\int_{0}^{t}d\tau e^{-\kappa \tau /2}\mathrm{Tr}%
_{r}\{[H_{I}(t),[H_{I}(t\mathtt{-}\tau ),\varrho (t)\otimes \rho
_{eq}]]\}.
\end{align}
Here, we have used the Eqs. \eqref{DCA} and \eqref{DCAP} to remove the interaction superoperator.
To calculate the above twice commutators, we define a function as follows:
\begin{equation}
\mathcal{F}^{\vec{n}}_{\vec{m}}(t,s)=\mathrm{Tr}_{r}[H^{n}_{m
}(t),[H^{n'}_{m'}(s),\varrho (t)\otimes \rho
_{eq}]],
\end{equation}
with
\begin{align}
  H^{n}_{m}(t)=g^{m}_{n}a_{m}^{+}\sigma _{-}^{n(\mathbf{z})}e^{i\Delta t}+H.c.,
\end{align}
where $\vec{n}=(n,n'),\vec{m}=(m,m')$.
Hence, the 2nd order TCL master equation (\ref{B11}) becomes
\begin{equation}
\dot{\varrho} (t)=-\sum_{ \vec{n}\vec{m}}\int_{0}^{t}d\tau e^{-\kappa
\tau /2}\mathcal{F}^{\vec{m}}_{\vec{n}}(t,t-\tau ).
\end{equation}%
We note that the resonators equilibrium
state satisfies
\begin{align}
&\mathrm{Tr}_{r}[a_{m}a_{m'}^{+}\rho _{eq}] =(\bar{n}%
+1)\delta^{m}_{m'}, \\
 &\mathrm{Tr}_{r}[a_{m}^{+}a_{m ^{\prime }}\rho _{eq}]=\bar{n}
\delta^{m}_{m'}, \\
&\mathrm{Tr}_{r}[a_{m }^{+}a_{m ^{\prime }}^{+}\rho _{eq}] =0, \\
&\mathrm{Tr}_{r}[a_{m }a_{m ^{\prime }}\rho _{eq}] =0.
\end{align}%
Thus, we reach
\begin{align}
\mathcal{F}^{\vec{m}}_{\vec{n}}&(t,s) = g^m_{n} g_{n'}^{m'\ast} \mathrm{Tr}_{r} \{[a_{m}^{+}\sigma_{-}^{n}(t),
  [a_{m'}\sigma_{+}^{n'}(s),\varrho (t)\otimes \rho _{eq}]]\} \notag \\
  & + g_{n}^{m\ast }g^{m'}_{n'} \mathrm{Tr}_{r}\{[a_{m }\sigma_{+}^{n}(t),
  [a_{m'}^{+}\sigma _{-}^{n'}(s),\varrho (t)\otimes \rho _{eq}]]\}\notag \\
  &= \delta^{m}_{m'}\left\{
  g^{m}_{n}g _{n'}^{m \ast }(\bar{n}+1)
  [\varrho \sigma _{+}^{n^{\prime }}(s)\sigma _{-}^{n}(t)-\sigma _{-}^{n}(t)\varrho \sigma _{+}^{n^{\prime }}(s)]\right.\notag\\
  &+g^{m}_{n}g_{n'}^{m\ast }\bar{n}
  [\sigma_{-}^{n}(t)\sigma _{+}^{n^{\prime }}(s)\varrho -\sigma _{+}^{n^{\prime }}(s)\varrho \sigma _{-}^{n}(t)]\notag\\
  & + g _{ n}^{m\ast }g ^{m}_{ n'}(\bar{n}+1)[\sigma _{+}^{n}(t)
  \sigma _{-}^{n^{\prime }}(s)\varrho-\sigma _{-}^{n^{\prime }}(s)\varrho \sigma _{+}^{n}(t) ] \notag\\
  &+ \left.g_{ n}^{m\ast }g^{m} _{n'}\bar{n}[\varrho \sigma _{-}^{n^{\prime }}(s)
  \sigma_{+}^{n}(t)-\sigma _{+}^{n}(t)\varrho \sigma _{-}^{n^{\prime }}(s) ] \right\}.
\end{align}
Here, we have included the time dependence of $e^{\pm i\Delta t}$ in Pauli matrix $\sigma _{\pm}^{n^{\prime }}(t)=\sigma _{\pm}^{n^{\prime }}e^{\pm i\Delta t}$.
We introduce the superoperator generators
\begin{align}
\mathcal{G}^{\vec{m}}_{\vec{n}}(t) \varrho(t)&= \int_0^\infty d\tau e^{-\kappa\tau/2} \mathcal{F}^{\vec{m}}_{\vec{n}}(t,t-\tau),
\end{align}
 the master equation reduces into
\begin{equation}
\frac{d}{dt}\varrho(t)=-\sum_{\vec{n}\vec{m}}\mathcal{G}^{\vec{m}}_{\vec{n}}(t) \varrho(t).
\end{equation}
Here, we have set the upper limit of the integral to infinity: $\int_0^t d\tau\rightarrow \int_0^\infty d\tau$. Let us calculate it term by term
\begin{align} \label{ahgaghfngu}
\mathcal{G}^{\vec{m}}_{\vec{n}}\varrho &= \delta^{m}_{m'}\int_0^\infty d\tau e^{-\kappa\tau/2+i\Delta\tau} \left\{ g^{m}_{n}
  g^{m*}_{n'}\bar{n}[\sigma^{n}_-\sigma_+^{n'}\varrho-\sigma^{n'}_+\varrho\sigma^{n}_{-}]\right.\notag \\
  &+ \left. g^m_{n}g^{m*}_{n'}(\bar{n}+1)
  [\varrho\sigma^{n'}_{+}\sigma^{n}_--\sigma^{n}_-\varrho\sigma^{n'}_+]\right\}\notag \\
  &+ \delta^{m}_{m'}\int_0^\infty d\tau e^{-\kappa\tau/2-i\Delta\tau}\left\{g^{m*}_{ n}g^{m}_{ n'}\bar{n}
  [\varrho\sigma^{n'}_{-}\sigma^{n}_+-\sigma^{n}_+\varrho\sigma^{n'}_-] \right.
  \notag \\
  &+ \left.g^{m*}_{n}g^{m}_{n'}(\bar{n}+1)
  [\sigma^{n}_+\sigma_-^{n'}\varrho-\sigma^{n'}_-\varrho\sigma^{n}_{+}]\right\}.
\end{align}
Using the following formula of integration
\begin{equation}
\int_0^\infty d\tau e^{-\kappa\tau/2}e^{\pm i\tau\Delta}=\frac{2}{\kappa \mp i2\Delta}=\eta\pm i\lambda,
\end{equation}
with
\begin{equation} \label{B19}
\eta=\frac{2\kappa}{\kappa^2+4\Delta^2}, \lambda=%
\frac{4\Delta}{\kappa^2+4\Delta^2},
\end{equation}
Eq. \eqref{ahgaghfngu} reduces into
\begin{align}
  \mathcal{G}_{\vec{n}\vec{m}}\varrho =\delta^{m}_{m'} \mathcal{G}^{m}_{\vec{n}}\varrho,
\end{align}
with
\begin{align} \label{vahvhargv}
\mathcal{G}^{m}_{\vec{n}}\varrho
  &=(\eta+ i\lambda)g^m_{ n}g^{m*}_{ n'}\bar{n}
  [\sigma^{n}_-\sigma_+^{n'}\varrho-\sigma^{n'}_+\varrho\sigma^{n}_{-}] \\
  & + (\eta- i\lambda)g^{m*}_{ n}g^m_{n'}\bar{n}
  [\varrho\sigma^{n'}_{-}\sigma^{n}_+-\sigma^{n}_+\varrho\sigma^{n'}_-] \notag \\
  &+ (\eta+ i\lambda)g^m_{ n}g^{m*}_{n'}(\bar{n}+1)
  [\varrho\sigma^{n'}_{+}\sigma^{n}_--\sigma^{n}_-\varrho\sigma^{n'}_+] \notag \\
  &+ (\eta- i\lambda)g^{m*}_{ n}g^m_{ n'}(\bar{n}+1)
  [\sigma^{n}_+\sigma_-^{n'}\varrho-\sigma^{n'}_-\varrho\sigma^{n}_{+}]. \notag
\end{align}
For simplicity, we exchange index $n$ and $n'$ in the second and fourth lines of Eq. \eqref{vahvhargv}, which leads into
\begin{align} \label{uncffvbggsl}
\sum_{\vec{n}}\mathcal{G}^{m}_{\vec{n}}\varrho
  &=\sum_{\vec{n}} g^m_{n}g^{m*}_{ n'}\left\{(\eta+ i\lambda)\bar{n}
  [\sigma^{n}_-\sigma_+^{n'}\varrho-\sigma^{n'}_+\varrho\sigma^{n}_{-}]\right. \notag \\
  & + (\eta- i\lambda)\bar{n}
  [\varrho\sigma^{n}_{-}\sigma^{n'}_+-\sigma^{n'}_+\varrho\sigma^{n}_-] \notag \\
  &+ (\eta+ i\lambda)(\bar{n}+1)
  [\varrho\sigma^{n'}_{+}\sigma^{n}_--\sigma^{n}_-\varrho\sigma^{n'}_+] \notag \\
  &+\left. (\eta- i\lambda)(\bar{n}+1)
  [\sigma^{n'}_+\sigma_-^{n}\varrho-\sigma^{n}_-\varrho\sigma^{n'}_{+}]\right\}.
\end{align}
Hence, one can divide Eq. \eqref{uncffvbggsl} into two parts
\begin{align}
\sum_{\vec{n}m}\mathcal{G}^{m}_{\vec{n}}\varrho
  &=\eta \mathcal{G}_{\eta}\varrho+i\lambda \mathcal{G}_{\lambda}\varrho,
\end{align}
with
\begin{align}
\mathcal{G}_{\eta}\varrho
  &=\sum_{\vec{n}m} g^m_{n}g^{m*}_{n'}\left[\bar{n}
  (\sigma^{n}_-\sigma_+^{n'}\varrho+\varrho\sigma^{n}_{-}\sigma^{n'}_+-2\sigma^{n'}_+\varrho\sigma^{n}_{-})\right. \notag \\
  &+ \left.(\bar{n}+1)
  (\varrho\sigma^{n'}_{+}\sigma^{n}_-+\sigma^{n'}_+\sigma_-^{n}\varrho-2\sigma^{n}_-\varrho\sigma^{n'}_{+})\right],
\end{align}
\begin{align}
\mathcal{G}_{\lambda}\varrho
  &=\sum_{\vec{n}m} g^{m}_{n}g^{m*}_{n'}\left[\bar{n}
  (\sigma^{n}_-\sigma_+^{n'}\varrho-\varrho\sigma^{n}_{-}\sigma^{n'}_+)\right. \\
  &+ \left.(\bar{n}+1)
  (\varrho\sigma^{n'}_{+}\sigma^{n}_-- \sigma^{n'}_+\sigma_-^{n}\varrho) \right].\notag
\end{align}

Next, let us calculate the expectation value of the projection operator $\mathcal{P}_a(\mathcal{E})=\vert \mathcal{E}\rangle _{a}\langle \mathcal{E}\vert$ at an arbitrary time $t$. For the $a$th qubit, it is defined as \begin{align}
  P^a_{\mathcal{E}}(t)=\textrm{Tr}_{q} \{\varrho(t) \mathcal{P}_a(\mathcal{E})\},
\end{align}
where $\textrm{Tr}_{q}=\textrm{tr}^1_{q}\cdots \textrm{tr}^N_{q}$ means trace over the multiqubit space. The time evolution of that reads
\begin{align} \label{B24}
\frac{d}{dt}& P^{a}_{\mathcal{E}}(t)=- \eta G^{a}_{\eta}(\mathcal{E})\varrho-i\lambda G^{a}_{\lambda}(\mathcal{E})\varrho,
\end{align}
with
\begin{align}
    G^a_{\eta/\lambda}(\mathcal{E})\varrho =\textrm{Tr}_{q}\{ \mathcal{G}_{\eta/\lambda}\varrho\mathcal{P}_a(\mathcal{E}) \},
\end{align}
i.e.,
\begin{align}
    G^a_{\eta}(\mathcal{E})&\varrho =\sum_{\vec{n}m} g^m_{ n}g^{m*}_{ n'}\textrm{Tr}_{q}\left \{ \varrho\left[\bar{n}
      (\mathcal{P}_a(\mathcal{E})\sigma^{n}_-\sigma_+^{n'}\right.\right.  \\
      &+\sigma^{n}_{-}\sigma^{n'}_+\mathcal{P}_a(\mathcal{E})-2\sigma^{n}_{-}\mathcal{P}_a(\mathcal{E})\sigma^{n'}_+)\notag  \\
      &+(\bar{n}+1)
      (\sigma^{n'}_{+}\sigma^{n}_-\mathcal{P}_a(\mathcal{E})+\mathcal{P}_a(\mathcal{E})\sigma^{n'}_+\sigma_-^{n}      \notag\\
    &\left. \left. -2 \sigma^{n'}_{+}\mathcal{P}_a(\mathcal{E})\sigma^{n}_-)\right] \right\}, \notag
\end{align}
\begin{align}
    G^a_{\lambda}(\mathcal{E})& \varrho =\sum_{\vec{n}m} g^{m}_{n}g^{m*}_{ n'}\textrm{Tr}_{q}\left \{ \varrho\left[\bar{n}
      (\mathcal{P}_a(\mathcal{E})\sigma^{n}_-\sigma_+^{n'}-\sigma^{n}_{-}\sigma^{n'}_+\mathcal{P}_a(\mathcal{E}))\right. \right.\notag  \\
    &+ \left. \left.(\bar{n}+1)
      (\sigma^{n'}_{+}\sigma^{n}_-\mathcal{P}_a(\mathcal{E})- \mathcal{P}_a(\mathcal{E})\sigma^{n'}_+\sigma_-^{n}) \right] \right \},
\end{align}
Note that i) for three different index ($a,n,n'$), $\mathcal{P}_a(\mathcal{E}), \sigma^{n}_-$ and $\sigma_+^{n'}$ commute, it is easy to see that these terms cancel with each other; ii) for the case of $n=n'\neq a$, one can quickly check that these terms also cancel with each other. Hence, we can divide the rest into three parts:
\begin{align}
  G^a_{\eta/\lambda}(\mathcal{E})\varrho = G^{a,11}_{\eta/\lambda}(\mathcal{E})\varrho+G^{a,10}_{\eta/\lambda}(\mathcal{E})\varrho+G^{a,01}_{\eta/\lambda}(\mathcal{E})\varrho,
\end{align}
with
\begin{align}
  G^{a,11}_{\eta/\lambda}(\mathcal{E})\varrho=\left.G^{a}_{\eta/\lambda}(\mathcal{E})\varrho\right\vert_{n=a,n'=a},
\end{align}
\begin{align}
  G^{a,10}_{\eta/\lambda}(\mathcal{E})\varrho=\left.G^{a}_{\eta/\lambda}(\mathcal{E})\varrho\right\vert_{n=a,n'\neq a},
\end{align}
\begin{align}
  G^{a,01}_{\eta/\lambda}(\mathcal{E})\varrho=\left.G^{a}_{\eta/\lambda}(\mathcal{E})\varrho\right\vert_{n\neq a,n'=a}.
\end{align}
Let us begin with simpler ones
\begin{align} \label{aguraignvn11}
  G^{a,11}_{\lambda}(\mathcal{E})\varrho &=\sum_{m} g^m_{ a}g^{m*}_{ a}\textrm{Tr}_{q}\left\{\varrho[\bar{n}
  (\delta^{-1}_{\mathcal{E}} \sigma^{a}_-\sigma_+^{a}-\sigma^{a}_{-}\sigma^{a}_+\delta^{-1}_{\mathcal{E}})\right. \notag \\
  &+ \left.(\bar{n}+1)
  (\sigma^{a}_{+}\sigma^{a}_-\delta^{+1}_{\mathcal{E}}- \delta^{+1}_{\mathcal{E}}\sigma^{a}_+\sigma_-^{a}) ] \right\}\notag\\
  &=0,
\end{align}
\begin{align} \label{aguraignvn10}
  G^{a,10}_{\lambda}(\mathcal{E})\varrho &=\sum_{m,n}^{n\neq a}g^m_{ a}g^{m*}_{ n}\textrm{Tr}_{q}\left\{\varrho[\bar{n}
  (\delta^{-1}_{\mathcal{E}} \sigma^{a}_-\sigma_+^{n}-\sigma^{a}_{-}\delta^{+1}_{\mathcal{E}}\sigma^{n}_+)\right. \notag \\
  &+ \left.(\bar{n}+1)
  (\sigma^{a}_-\delta^{+1}_{\mathcal{E}}\sigma^{n}_{+}- \delta^{-1}_{\mathcal{E}}\sigma_-^{a} \sigma^{n}_+) ] \right\}\notag\\
  &=+\mathcal{E}\sum_{n\neq a} C_{n a} \langle \sigma^{n}_{+}\sigma^{a}_{-}\rangle (t),
\end{align}
\begin{align} \label{aguraignvn01}
  G^{a,01}_{\lambda}(\mathcal{E})\varrho &=\sum_{m,n}^{n\neq a} g^m_{ n}g^{m*}_{ a}\textrm{Tr}_{q}\left\{\varrho[\bar{n}
  (\delta^{+1}_{\mathcal{E}} \sigma_+^{a}\sigma^{n}_--\sigma^{a}_+\delta^{-1}_{\mathcal{E}}\sigma^{n}_{-})\right. \notag \\
  &+ \left.(\bar{n}+1)
  (\sigma^{a}_{+}\delta^{-1}_{\mathcal{E}}\sigma^{n}_-- \delta^{+1}_{\mathcal{E}} \sigma^{a}_+\sigma_-^{n}) ] \right\}\notag\\
  &= -\mathcal{E}\sum_{n\neq a} C^*_{n a}\langle \sigma^{n}_{-}\sigma^{a}_{+}\rangle (t),
\end{align}
with
\begin{equation} \label{B29}
C _{na}= \sum_{m}g^m_{ a}g^{m*}_{ n}.
\end{equation}%
Here, we define the qubit-qubit correlation functions
\begin{align}
  \langle \sigma^{n}_{\alpha}\sigma^{n'}_{\alpha'}\rangle (t)=\textrm{Tr}_{q} \{\varrho(t) \sigma^{n}_{\alpha}\sigma^{n'}_{\alpha'}\}.
\end{align}
Hence, we obtain
\begin{align} \label{agurajgagnvn}
  G^a_{\lambda}(\mathcal{E})\varrho &=\mathcal{E}\sum_{n\neq a} [C_{n a}\langle \sigma^{n}_{+}\sigma^{a}_{-}\rangle (t) - C^*_{n a}\langle \sigma^{n}_{-}\sigma^{a}_{+}\rangle (t)].
\end{align}
Next, we calculate another term
\begin{align}
  G^{a,11}_{\eta}(\mathcal{E})\varrho &=2\sum_{m} \vert g^m_{ a}\vert ^2 \textrm{tr}_{q}\{\varrho \left[\bar{n}
  ( \delta^{-1}_{\mathcal{E}} \sigma^{a}_-\sigma_+^{a}-\sigma^{a}_{-}\delta^{+1}_{\mathcal{E}}\sigma^{a}_+) \right. \notag \\
  &+\left. (\bar{n}+1)
  (\sigma^{a}_{+}\sigma^{a}_- \delta^{+1}_{\mathcal{E}}- \sigma^{a}_{+}\delta^{-1}_{\mathcal{E}} \sigma^{a}_- )\right] \}\notag\\
  &=2\mathcal{E}C_{aa} [(\bar{n}+1)P^{a}_{+1}(t)-\bar{n}P^{a}_{-1}(t)],
\end{align}
\begin{align}\label{ahhgrgvhua2}
  G^{a,10}_{\eta}(\mathcal{E})\varrho &= \sum_{m,n}^{n\neq a} g^m_{a}g^{m*}_{ n}\textrm{tr}_{q}\left\{\varrho\left[\bar{n}
  (\delta^{-1}_{\mathcal{E}}\sigma^{a}_-\sigma_+^{n} -\sigma^{a}_{-}\delta^{+1}_{\mathcal{E}}\sigma^{n}_+) \right.\right. \notag \\
  &+(\bar{n}+1)
  (\sigma^{a}_-\delta^{+1}_{\mathcal{E}}\sigma^{n}_{+}- \left.\left. \delta^{-1}_{\mathcal{E}}\sigma^{a}_-\sigma^{n}_{+})\right] \right\} \notag\\
  &= \mathcal{E} \sum_{n\neq a} C_{n a}\langle \sigma^{n}_{+}\sigma^{a}_{-}\rangle (t),
\end{align}
\begin{align}\label{ahhgrgvhua1}
  G^{a,01}_{\eta}(\mathcal{E})\varrho &= \sum_{m,n}^{n\neq a} g^m_{ n}g^{m*}_{ a}\textrm{tr}_{q}\left\{\varrho\left[\bar{n}
  (\sigma^{a}_+\delta^{-1}_{\mathcal{E}}\sigma^{n}_{-}-\delta^{+1}_{\mathcal{E}}\sigma^{a}_+\sigma^{n}_{-}) \right.\right. \notag \\
  &+\left.\left.(\bar{n}+1)
  (\delta^{+1}_{\mathcal{E}} \sigma^{a}_+\sigma_-^{n} - \sigma^{a}_{+}\delta^{-1}_{\mathcal{E}}\sigma^{n}_-)\right] \right\} \notag \\
  &=\mathcal{E} \sum_{n\neq a} C^*_{n a}\langle \sigma^{n}_{-}\sigma^{a}_{+}\rangle (t).
\end{align}
Hence, we reach
\begin{align}
  G^{a}_{\eta}(\mathcal{E})\varrho &=2\mathcal{E}C_{aa} [(\bar{n}+1)P^{a}_{+1}(t)-\bar{n}P^{a}_{-1}(t)]\notag\\
  &+ \mathcal{E} \sum_{n\neq a} [C_{n a}\langle \sigma^{n}_{+}\sigma^{a}_{-}\rangle (t)+ C^*_{n a}\langle \sigma^{n}_{-}\sigma^{a}_{+}\rangle (t)].
\end{align}

\begin{figure}[t]
\centering
\includegraphics[height=0.6 \columnwidth]{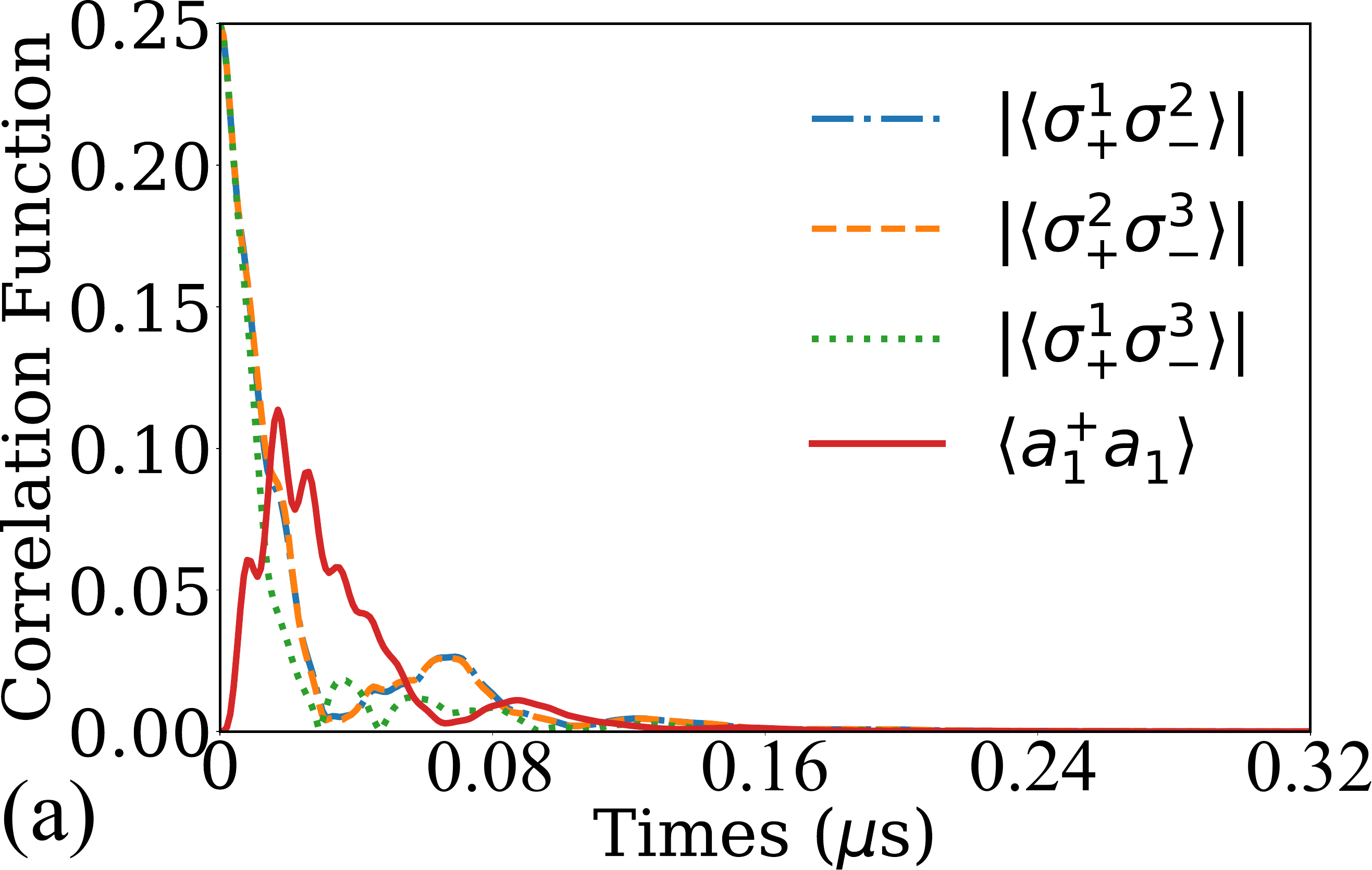}
\includegraphics[height=0.6 \columnwidth]{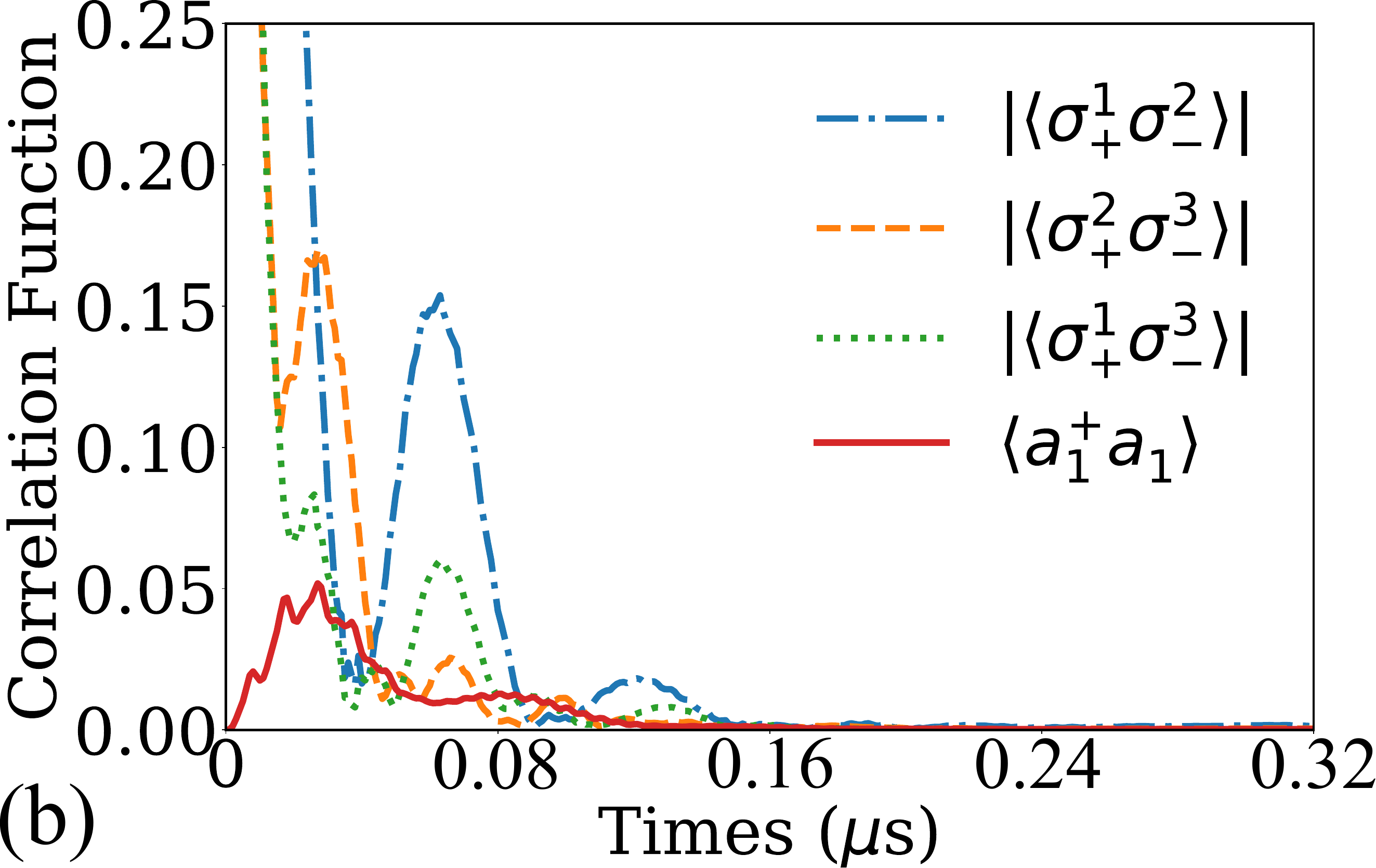}
\caption{The evolution of the simulated correlation functions $\langle
\sigma^n_+\sigma^a_- \rangle$ and $\langle
a^{+}_1a^{}_1 \rangle$ for three qubits with different target state (a) $[\sigma^1_{\mathbf{z}},\sigma^2_{\mathbf{z}},\sigma^3_{\mathbf{z}}]=[\sigma_{z},\sigma_{z},\sigma_{z}]$ and (b) $[\sigma^1_{\mathbf{z}},\sigma^2_{\mathbf{z}},\sigma^3_{\mathbf{z}}]=[\sigma_{x},\sigma_{y},\sigma_{z}]$.
Here, $(g,\protect\kappa)/2\protect\pi%
=(15,10)$ MHz, and qubit dissipation rates are set to be zero, ($1/T_\protect%
\protect\theta$, $1/T_\protect\protect\phi$)=(0,0). Other, parameters are the same as Fig. \ref{f2}, and the temperature is zero, i.e., $T_c=0$K.
}
\label{f4}
\end{figure}

With the definition of $\vec{P}^a(t)=[P^a_{-1}(t),P^a_{+1}(t)]^{T}$ and $\vec{Q}^{na}(t)=[\langle \sigma^{n}_{+}\sigma^{a}_{-}\rangle (t),\langle \sigma^{n}_{-}\sigma^{a}_{+}\rangle (t)]^T$, the rate equation (\ref{B24}) reduces
to a rate equation for the state populations:
\begin{equation} \label{fvjkvk}
\frac{d}{dt}\vec{P}^{a}(t)=-\Gamma _{a}\mathbf{N}%
\vec{P}^{a}(t)-\sum_{n\neq a }\vec{\Gamma}_{na}\vec{Q}^{na}(t),
\end{equation}%
with
\begin{equation}
\mathbf{N}=\left[
\begin{array}{cc}
\bar{n} & -(\bar{n}+1) \\
-\bar{n} & +(\bar{n}+1)%
\end{array}%
\right],
\end{equation}%
\begin{equation}
\vec{\Gamma}_{na}=\left[
\begin{array}{cc}
- \Gamma _{na} & - \Gamma^* _{na} \\
+ \Gamma _{na} & + \Gamma^* _{na}
\end{array}%
\right].
\end{equation}%
The effective dissipation rate reads
\begin{equation} \label{B27}
\Gamma _{a}=2\eta C_{aa}=\frac{(1+\cos \theta _{a})^{2}}{1+4(\Delta /\kappa )^{2}}%
\frac{g^{2}}{\kappa },
\end{equation}
\begin{align}
\Gamma _{na}&=(\eta +i\lambda) C _{na}=-(\eta +i\lambda)\frac{g^2}{8}e^{i(\phi _{a}-\phi _{n})} \notag \\
&\times (\cos \theta _{a}+ 1)(\cos \theta _{n}+ 1),
\end{align}
For the case of $\Delta=0$ ($\lambda=0$), we obtain
\begin{equation}
\Gamma _{a}=\frac{16}{\kappa}\vert g_{a}\vert ^2 =\frac{g^{2}}{\kappa }(1+\cos \theta _{a})^{2}
\end{equation}
\begin{align}
\Gamma _{na}&=-\frac{4}{\kappa}g_ng_ae^{i(\phi _{a}-\phi _{n})} =-\frac{g^2}{4\kappa }e^{i(\phi _{a}-\phi _{n})} \notag \\
&\times (\cos \theta _{a}+ 1)(\cos \theta _{n}+ 1).
\end{align}
$\Gamma _{n'n}$ describes the correction of dynamics caused by the nonlocal qubit-qubit correlation, which results from the indirect qibit-qubit interaction realized by photon-assisted qubit-filter coupling.
However, this nonlocal qubit-qubit correlation will be quickly consumed by dissipation caused by vast photon loss of resonators, as shown in Fig. \ref{f4}.
Note that there is no nonlocal qubit-qubit correlation i.e., $ \langle \sigma^{n}_{\pm}\sigma^{a}_{\mp}\rangle (t\rightarrow \infty )=0$, and rate equation \eqref{fvjkvk} reduces into
\begin{equation}
\frac{d}{dt}\vec{P}^{a}(t)=-\Gamma _{a}\mathbf{N}%
\vec{P}^{a}(t)\;.
\end{equation}%

In addition to the quantitative calculation of the reset time mentioned above, we present a qualitative discussion. This procedure's required time depends on the loss speed of resonators' photon, which is the only dissipative channel relative to each resonator's dissipation rate and photon population. The latter is plotted by the red lines in Fig. \ref{f4} and relates the reset time to effective qubit-filter coupling $g_n$. Unprejudiced indirect photon exchanging,
realized by the effective qubit-filter coupling, only uniformizes each resonator's photon population of
making an ignorable difference in the photon population
in an ideal case with the same $g_n$. In our circuit QED architecture, the more
qubits, the more resonators, which causes an equal photon sharing population on each qubit. Thus, the required reset time is independent
of the number of qubits.
Finally, we emphasize that $H_I$ will
maintain even in a 2D or 3D circuit network. Therefore, our discussion will
still be available in 2D and 3D cases.

%

%\bibliographystyle{unsrt}
%\bibliography{reference}

\end{document}